\documentstyle[12pt]{article}
\normalbaselines
\catcode `\@=11
\@addtoreset{equation}{section}
 
\def\section{\@startsection {section}{1}{\z@}{-3.5ex plus -1ex minus
     -.2ex}{2.3ex plus .2ex}{\normalsize\bf}}
\def\subsection{\@startsection{subsection}{2}{\z@}{-3.25ex plus -1ex minus
 -.2ex}{1.5ex plus .2ex}{\normalsize\bf}}

\def\@cite#1#2{${}^{\mbox{\scriptsize#1\if@tempswa , #2\fi}}$}
\def\thebibliography#1{\section*{References\markboth
  {REFERENCES}{REFERENCES}}\list
  {\arabic{enumi}.}{\settowidth\labelwidth{[#1]}\leftmargin\labelwidth
  \advance\leftmargin\labelsep
  \usecounter{enumi}}
  \def\newblock{\hskip .11em plus .33em minus -.07em}
  \sloppy
  \sfcode`\.=1000\relax}
 
\catcode `\@=12
\begin{document}
{\bf DIRECTLY INTERACTING MASSLESS PARTICLES - 
A TWISTOR APPROACH \footnote{In November 1995 accepted for publication in JMP.}.}\vspace{5.3cm}\\
\noindent
\hspace*{1in}
\begin{minipage}{13cm} Andreas Bette, \vspace{0.1cm}\\
Stockholm University, \vspace{0.1cm}\\
Department of Physics, \vspace{0.1cm}\\
Box 6730, \vspace{0.1cm}\\
S-113 85 Stockholm, Sweden.\vspace{0.1cm}\\
tfx: +468-347817 att. Andreas Bette, \vspace{0.1cm}\\
e-mail: $<$ab@vanosf.physto.se$>$
\end{minipage}

\vspace*{0.5cm}

\begin{abstract}

\noindent 
Twistor phase spaces are used to provide a general description of the
dynamics of a finite number of directly interacting massless spinning
particles forming a closed relativistic massive and spinning system
with an internal structure.

\vskip 8pt 

\noindent 
A Poincar{\'e} invariant canonical quantization of the so obtained
twistor phase space dynamics is performed.

\end{abstract}

\vfill
\eject

\section{\hspace{-4mm}\hspace{2mm}INTRODUCTION.} 

It is possible that "elementary" massive particles such as electron,
proton, neutron etc.  should be regarded as bound states of a
{\underline {finite}} number of massless and spinning interacting
parts.

\vskip 10pt

\noindent
In order to investigate such a possibility we develop in this paper a
general formalism with its roots in the Twistor Theory of
Penrose\cite{prmc} and in the Theory of Action at a Distance in
Relativistic Particle Dynamics (in its instantaneous form)\cite{wfw}.

\vskip 10pt

\noindent
Somewhat similar attempts to classify elementary
particles employing the Twistor Theory but without any explicit
mention of the Theory of Action at a Distance in Relativistic Particle
Dynamics (R.a-a-a-d), have been made before by Hughston\cite{hl} and 
Popovich\cite{pa}.  Certain other developments in the same direction
appeared in papers written by Perj{\`e}s et.  al.\cite{pz,pzl} and
Sparling\cite{spar}.

\vskip 10pt

\noindent
The quantum version of R.a-a-a-d in connection with the Twistor Theory
seems to be implicit in an example worked out by Hughston\cite{hl1}.

\vskip 10pt

\noindent 
The framework we are presenting is however from the very beginning
completely in accordance with R.a-a-a-d.  Classical states, which
correspond to relativistic quantum bound states, of a massive and
spinning composite particle, are represented by points in a finite
dimensional "twistor phase space".  The description is purely
hamiltonian.  The suggested quantization procedure is simultaneously
canonical and Poincar{\'e} invariant.  The arising canonically
conjugated quantum mechanical operators represent "square roots" of
the null-momenta and "square roots" of the "positions" attributed to
(the classical limit of) the massless constituents forming a massive
system.

\vskip 10pt

\noindent 
Exploring the idea of instantaneous relativistic action at a
distance\cite{wfw} in the phase space of two twistors we have shown
previously\cite{ab1} how a free massive, spinning point-like particle
may be thought of as a relativistic rigid rotator (endowed with
intrinsic spin) composed of two massless spinning parts.  The term
"instantaneous" refers to the rest frame defined by the total
time-like four-momentum of the rigid rotator itself.  The present
paper may be regarded as an extension and generalisation of the same
idea.  In its very rough state the idea appeared in our 
report\cite{bette} from 1979.

\vskip 10pt

\noindent 
The work is organized as follows: First in the next section some
results from the Twistor Theory are reformulated in a way which
exhibits how they tie in with a relativistic, classical, finite
dimensional phase space mechanics of a massless particle with
helicity.  Ten Poincar{\'e} covariant functions fulfilling Poisson
bracket algebra of the Poincar{\'e} group and a function corresponding
to the helicity operator are identified\cite{prmc}.

\vskip 10pt

\noindent 
These well-known results in new clothes are generalised in section
three where a twistor phase space of a massive spinning system
composed of an arbitrary finite number of massless parts is
introduced.

\vskip 10pt

\noindent 
Again ten Poincar{\'e} covariant functions fulfilling Poisson
bracket algebra of the Poinca- r{\'e} group are identified.  
Four functions representing the (real) position
four-vector of the total massive spinning system 
are also identified\cite{hl}.  

\vskip 10pt

\noindent 
A fundamental set of Poincar{\'e} scalar functions forming a closed
finite Poisson bracket subalgebra is identified.  These functions
serve as generators (eigenvalues of their quantum mechanical
counterparts are tentatively identified with such quantum numbers as
electric charge, baryonnumber etc.) of the internal symmetries.

\vskip 10pt

\noindent 
A general formula for the four functions representing the
Pauli-Luba{\'n}ski spin four-vector is derived.

\vskip 10pt

\noindent 
A general formula for the function representing the square of the
relativistic spin in terms of the generators of the internal
symmetries is found.

\vskip 10pt

\noindent 
It is discovered that, provided the relativistic spin does not vanish,
the components of the position four-vector of the total system do not
Poisson commute (see (3.27)).  

\vskip 10pt

\noindent 
A certain class of Poincar{\'e} scalar functions in the twistor phase
space is selected. Functions in this class, when used as
generators of motion, produce canonical flows which in Minkowski space
describe a finite number of mutually interacting spinning massless
particles forming a closed freely moving massive and, in general,
spinning relativistic system.

\vskip 10pt

\noindent
The so obtained relativistic dynamics constitutes our general
dynamical principle. 

\vskip 10pt

\noindent 
In the last section this classical phase space dynamics is
canonically quantized in a way which corresponds to the real
polarization of the twistor phase space\cite{woo}.  A Poincar{\'e}
invariant scalar product, on the space of functions representing
quantum states of the massive spinning system composed of a finite
number of massless parts, is introduced.

\vskip 10pt

\noindent
The following notation will be used:

\vskip 10pt

\noindent 
Latin letters with lower case latin indices will denote four-vectors
and four-tensors.  Lower case greek letters with upper case latin
indices (either primed or unprimed) will denote spinors.  Upper case
latin letters with lower case greek indices will denote non-projective
twistors.  Lower case latin indices within round brackets are used to
number the different massless parts and in this way label the internal
degrees of freedom.  A bar over a letter or over an expression denotes
complex conjugation.  The usual summation convention over repeted
indices is assumed throughout.  The physical units are so chosen that
$c={\hbar}=1$.  The signature of the metric $g_{ij}$ in Minkowski
space is taken to be $+---$.  The fully antisymmetric alternating
four-tensor will be denoted by $\epsilon_{ijkl}$.

\section{\hspace{-4mm}\hspace{2mm}THE ELEMENTARY TWISTOR PHASE SPACE.} 

There are no new results in this section except from the way they are
presented.

\vskip 10pt

\noindent 
We introduce the notion of a twistor phase space \bf Tp \rm which
will be regarded as a space of classical states arising as a limit of
some corresponding (not yet specified) quantum mechanical description
of a massless particle with helicity.  The value of such a quantum
mechanical helicity is supposed to arise as an eigenvalue of some
appropriate (not yet specified) quantum mechanical helicity operator.
In the classical limit this quantum mechanical helicity operator
should then correspond to a real valued function on \bf Tp.  \rm
Therefore the classical helicity (being a limit of a quantum
mechanical operator whose eigenvalues are discrete) is not a discrete
variable.  In addition, following Penrose\cite{prmc} and
Hughston\cite{hl}, ten real valued Poincar{\'e} covariant functions
(corresponding to the generators of the Poincar{\'e} algebra) on \bf
Tp \rm are identified as (classical) physical observables.

\vskip 10pt

\noindent
\it DEFINTION 1: \rm

\vskip 10pt

\noindent
A non-projective twistor space \bf T \rm is a four dimensional complex
vector space (i.e.  $C^{4} \cong R^{8}$) endowed with the isometry
group $SU(2,2)$.

\vskip 10pt

\noindent
\sl REMARK 1: \rm

\vskip 10pt

\noindent
$SU(2,2)$ is to be identified with (the universal covering of) the so
called conformal group of the compactified Minkowski space and
it contains as one of its subgroups (the universal covering of) the
Poincar{\'e} group which, in turn, contains as one of its subgroups
(the universal covering of) the Lorentz group i.e.  $SL(2,C)$.  As
well-known the Poincar{\'e} group is the isometry group of the
physical Minkowski space.

\vskip 10pt

\noindent
In order to see how vectors in \bf T \rm are related to physical
Minkowskian quantities (such as angular and linear four-momenta,
position four-vectors and Poincar{\'e} invariant scalars) it is
convenient to choose a basis in \bf T \rm in a very special way.  A
vector in \bf T \rm given with respect to such a basis is called a
non-projective twistor.  With respect to any such a twistor basis the
$SU(2,2)$ metric is non-diagonal.

\vskip 10pt

\noindent
\it DEFINTION 2: \rm

\vskip 10pt

\noindent
A non-projective twistor $Z^{\alpha}$ and the corresponding
(twistor) complex conjugated twistor ${\bar Z}_{\alpha}$ may thus be
represented by two Weyl spinors and their conjugates:

\begin{equation}Z^{\alpha} = (\omega^{A},\ \pi_{A^\prime}),
\qquad \qquad 
{\bar  Z}_{\alpha} = ( {\bar  \pi_{A}},\ {\bar  \omega}^{A^\prime}).\end{equation}

\vskip 10pt

\noindent
\sl REMARK 2: \rm

\vskip 10pt

\noindent
Such a spinor representation of a non-projective twistor and its
twistor conjugate also explicitly shows how the Poincar{\'e} group
acts on \bf T.  \rm Coordinates of the two spinors represented by
$\pi_{A^\prime}$ and $\omega^{A}$ are covariant with respect to the
(identity connected part of the) Lorentz group while four-translations
$T^{a}$ act only on the "$\omega$" spinor parts of the twistor $Z$ and
its (twistor) complex conjugate\cite{prmc} $\bar Z.$

\vskip 10pt

\noindent
\it DEFINTION 3: \rm

\vskip 10pt

\noindent
The elementary twistor phase space \bf Tp \rm is spanned by $R^{8}$ in
which each point is labelled by a non-projective twistor and its
twistor complex conjugate.  Further, \bf Tp \rm is equipped with an
$SU(2,2)$ invariant symplectic structure\cite{hl,pr2,prrw} defined by
the following canonical Poisson bracket relations:

\begin{equation}
\{Z^{\alpha},\ {\bar  Z}_{\beta}\} = 
i{\delta}^{\alpha}_{\beta},
\qquad \qquad
\{Z^{\alpha},\ Z^{\beta}\} = 
\{{\bar  Z}_{\alpha},\ {\bar  Z}_{\beta}\} = 0,
\end{equation} 

\vskip 10pt

\noindent 
which, when written out in terms of spinors, reads:

\begin{equation}\{\omega^{A},\ {\bar  \pi_{B}}\} = 
i{\delta}^{A}_{B},
\qquad \qquad 
\{\pi_{B^\prime},\ {\bar  \omega}^{A^\prime}\} = 
i{\delta}^{A^\prime}_{B^\prime},\end{equation}

\begin{equation}\{\omega^{A},\ \omega^{B}\} = \{\omega^{A},\ \pi_{A^\prime}\}=
\{\pi_{A^\prime},\ \pi_{B^\prime}\} = 
\{\pi_{A^\prime},\ {\bar  \pi}_{B}\} = 0,\end{equation}

\begin{equation}\{{\bar  \omega}^{A^\prime},\ {\bar  \omega}^{B^\prime}\}=
\{{\bar  \omega}^{A^\prime},\ {\bar  \pi_{A}}\} = 
\{{\bar  \pi_{A}},\ {\bar  \pi_{B}}\} = 
\{{\omega}^{A},\ {\bar  \omega}^{B^\prime}\} = 0.\end{equation}

\vskip 10pt

\noindent
\sl REMARK 3: \rm

\vskip 10pt

\noindent
Points in \bf Tp \rm represent classical states of (the classical
limit of) a massless particle with (any value of its) helicity.

\vskip 10pt

\noindent
\it LEMMA 1: \rm

\vskip 10pt

\noindent
If the linear four-momentum $P_{a}$ and the angular four-momentum
${M}_{ab} = - { M}_{ba}$ of a massless particle (Penrose's abstract
index notation\cite{pr1} is used throughout the paper when
appropriate) are defined\cite{prmc} by the following set of
Poincar{\'e} covariant functions on \bf Tp: \rm

\begin{equation}P_{a}:= \pi_{A^\prime}{\bar  \pi}_A,\end{equation}

\begin{equation}
{  M}_{ab}:= i{\bar  \omega}_{(A^\prime}{\pi}_{B^{\prime})}
{\epsilon}_{AB}-i{\omega}_{(A}{\bar  \pi}_{B)} 
{\epsilon}_{A^{\prime}B^{\prime}},
\end{equation}

\vskip 10pt 

\noindent
then the canonical Poisson brackets (2.3)-(2.5)
imply that $P_{a}$
and $M_{ab}$ fulfil the Poisson bracket relations of the
Poincar{\'e} algebra\cite{hl}:

\begin{equation}\{P_{a},\ P_{b}\}=0,\end{equation}

\begin{equation}\{{  M}_{ab},\ P_{c}\}=2g_{c[a}P_{b]},\end{equation}

\begin{equation}\{{  M}_{ab},\ {  M}_{cd}\}=
2(g_{c[a}{  M}_{b]d}) + g_{d[b}{  M}_{a]c}).\end{equation}

\vskip 10pt

\noindent
Proof, which we omit, is just a tedious but straightforward computation.
Penrose's blob notation\cite{penr2} may be useful for this.

\vskip 10pt

\noindent
\sl REMARK 4: \rm

\vskip 10pt

\noindent
The above Poisson bracket relations define the momentum
mapping for the action of the Poincar{\'e} group on \bf Tp. \rm

\vskip 10pt

\noindent
\sl REMARK 5: \rm

\vskip 10pt

\noindent
A point in \bf Tp \rm carries more information about the
classical state of a massless particle than just information about its
linear and angular four-momenta.  It also defines its helicity and its
phase. 

\vskip 10pt

\noindent
\it LEMMA 2: \rm

\vskip 10pt

\noindent
The helicity (state) function is given by\cite{prmc}:

\begin{equation}s = {1 \over2}
(Z^{\alpha}\bar  Z_{\alpha}) =  
{1 \over2}
({\omega^{A}}{\bar  \pi_{A}} +
{\pi_{A^\prime}}{\bar  \omega^{A^\prime}}),
\end{equation}

\vskip 10pt

\noindent
which may be easily deduced if in the definition of the
Pauli-Luba{\'n}ski spin four-vector:

\begin{equation}
S^{a}:={1 \over 2}{\epsilon}^{abcd}P_{b}M_{cd}
\end{equation}

\vskip 10pt 

\noindent
the expressions in (2.6)-(2.7) and the spinor version of ${\epsilon}^{abcd}$
are inserted. The result in (2.11) follows from a simple spinor algebra
calculation\cite{prrw} which yields:

\begin{equation}S^{a}=sP^{a}.\end{equation}

\vskip 10pt

\noindent
\sl REMARK 6: \rm

\vskip 10pt

\noindent
Note that the massless particle's helicity function $s$ coincides
with one half of the $SU(2,2)$ norm of the corresponding
non-projective twistor.

\section{\hspace{-4mm}\hspace{2mm}THE GENERAL TWISTOR PHASE SPACE.} 

A generalisation of the results presented in the previous section
opens some new ways for applications of the Twistor Theory and of the
R.a.a.a.

\vskip 10pt

\noindent
Namely, it becomes possible to formulate a general dynamical principle
which, according to our interpretations and identifications, describes
a closed massive and, in general, spinning system composed of a finite
number of mutually interacting massless and spinning parts.

\vskip 10pt

\noindent
A direct product of any number of \bf Tp \rm may be used to define a
(reducible) phase space for a massive spinning relativistic particle
built up out of the massless ones.  In such a direct product \bf Tp(n)
\rm of $n$ ($n \geq 2$) copies of the elementary twistor phase space
\bf Tp \rm we exclude all points on all diagonals i.e.  each point in
\bf Tp(n) \rm represents a state of $n$ massless particles with their
four-momenta pointing at $n$ non-coinciding null-directions.

\vskip 10pt

\noindent
Generalising the definition $3$ of the previous section we let the
symplectic structure on the product twistor phase space \bf Tp(n) \rm
be given by the following set of canonical conformally invariant
Poisson brackets:

\vskip 10pt

\noindent
\it DEFINTION 4: \rm

\begin{equation}
\{Z^{\alpha}_{(i)},\ {\bar  Z}_{\beta (j)}\} = 
i{\delta}^{\alpha}_{\beta}{\delta}_{(i)(j)},
\ \
\{Z^{\alpha}_{(i)},\ Z^{\beta}_{(j)}\} = 
\{{\bar  Z}_{\alpha (i)},\ {\bar  Z}_{\beta (j)}\} = 0,
\ \   (i), (j) = 1,2..,n
\end{equation} 

\vskip 10pt

\noindent
where the index within brackets labels the $n$ distinct massless
parts.

\vskip 10pt

\noindent
\it LEMMA 3: \rm

\vskip 10pt

\noindent
If the linear and angular four-momenta functions of the massive and
spinning particle, formed by the $n$ massless spinning constituents,
are defined by:

\begin{equation} 
{\cal P}_{a}:={\pi_{A^{\prime} (i)}}{\bar \pi}_{A (i)},
\end{equation} 

\begin{equation} 
{\cal M}_{ab}:= 
i{\bar  \omega}_{(j)(A^\prime}{\pi}_{B^{\prime}) (j)}
{\epsilon}_{AB}-i{\omega}_{(j) (A}{\bar  \pi}_{B) (j)} 
{\epsilon}_{A^{\prime}B^{\prime}},
\end{equation} 

\vskip 10pt

\noindent
then the canonical commutation relations in (3.1) imply:

\begin{equation}
\{{\cal P}_{a},\ {\cal P}_{b}\}=0,
\end{equation}

\begin{equation}\{{\cal M}_{ab},\ {\cal P}_{c}\}=2g_{c[a}{\cal P}_{b]},\end{equation}

\begin{equation}\{{\cal M}_{ab},\ {\cal M}_{cd}\}=
2(g_{c[a}{\cal M}_{b]d}) + g_{d[b}{\cal M}_{a]c}),\end{equation}

\vskip 10pt 

\noindent
which, as should be expected, again represents Poisson bracket algebra of
the Poincar{\'e} group.

\vskip 10pt

\noindent
\sl REMARK 7:\rm

\vskip 10pt

\noindent
Using the canonical (conformally covariant) twistor coordinates (on
the $8n$ (real) dimensional twistor phase space) we note that they may
be used to form $2n^{2}-n$ real valued Poincar{\'e} scalar functions.
$n^{2}$ of these are also conformally (i.e.  $SU(2,2)$) invariant.

\vskip 10pt

\noindent
\it DEFINITION 5: \rm

\vskip 10pt

\noindent
The $n^{2}$ real valued $SU(2,2)$ invariant scalars are represented by
real and imaginary parts of the following functions:

\begin{equation}
a_{(i)(j)}:=Z^{\alpha}_{(i)}{\bar  Z}_{\alpha (j)},
\qquad \qquad
{\bar a}_{(i)(j)}=a_{(j)(i)} 
\end{equation}

\vskip 10pt 

\noindent
while the remaining Poincar{\'e} invariant scalars are represented by
real and imaginary parts of:

\begin{equation}
m_{(i)(j)} 
:=
I_{\alpha \beta}Z^{\alpha}_{(i)}Z^{\beta}_{(j)}=
\epsilon^{C^{\prime}D^{\prime}}
\pi_{D^{\prime}(i)}\pi_{C^{\prime} (j)}
=
-m_{(j)(i)}
\end{equation}

\begin{equation}
{\bar m}_{(i)(j)}
:=
I^{\alpha \beta}
{\bar  Z}_{\alpha (i)} {\bar  Z}_{\beta (j)}=
\epsilon^{CD}
{\bar \pi}_{D (i)}{\bar \pi}_{C (j)}
=
-{\bar m}_{(j)(i)}
\end{equation}

\vskip 10pt

\noindent 
where $\epsilon^{AB}$, $\epsilon^{A^{\prime}B^{\prime}}$ denote the
metric in the Weyl spinor space or equivalently $I^{\alpha \beta}$ and
$I_{\alpha \beta}$ denote the so called infinity twistor and its
dual\cite{prmc,prrw}.

\vskip 10pt

\noindent
\it LEMMA 4: \rm

\vskip 10pt

\noindent
From the canonical commutation relations in (3.1) it almost
trivially follows that:

\begin{equation}
\{a_{(i)(j)}, \ Z^{\alpha}_{(k)} 
\}=-i
\delta_{(j)(k)}Z^{\alpha}_{(i)}
\qquad \qquad
\{a_{(i)(j)}, \ {\bar Z}_{\alpha (k)} 
\}=i
\delta_{(i)(k)}{\bar Z}_{\alpha (j)}
\end{equation}

\begin{equation}
\{m_{(i)(j)}, \ Z^{\alpha}_{(k)} \}= 0
\qquad \qquad
\{{\bar m}_{(i)(j)}, \ Z^{\alpha}_{(k)}  \}=
2iI^{\alpha \mu}{\bar Z}_{\mu [(i)}\delta_{(j)](k)}	
\end{equation}

\begin{equation}
\{m_{(i)(j)}, \ {\bar Z}_{\alpha (k)}   \}=
2iI_{\mu \alpha}Z^{\mu}_{[(i)}\delta_{(j)](k)}	
\qquad \qquad
\{{\bar m}_{(i)(j)}, \ {\bar Z}_{\alpha (k)}   \}=0
\end{equation}

\vskip 10pt

\noindent
\it LEMMA 5: \rm

\vskip 10pt

\noindent
As shown by Hughston\cite{hl} the real four-vector valued function on
\bf Tp(n) \rm representing, in the Minkowski space, position
four-vector of the total system (composed of $n$ massless parts) is
given by:

\begin{equation} 
X^{a}=X^{A{A^\prime}}:={{\cal M}^{ab}{\cal P}_{b} \over m^{2}}
+{l \over m^{2}}{\cal P}^{a}
\end{equation} 

\vskip 10pt 
\noindent
where

\begin{equation} 
l:=-{1 \over 2}
(i\omega^{A}_{(i)}{\bar  \pi}_{A (i)} -
i\pi_{A^\prime (i)} {\bar  \omega}^{A^\prime}_{(i)}),
\end{equation} 

\vskip 10pt 

\noindent
and where

\begin{equation} 
m^{2}:={\cal P}_{a}{\cal P}^{a}=m_{(i)(j)}{\bar m}_{(i)(j)}
\end{equation} 

\vskip 10pt 

\noindent
or equivalently by\cite{hl}:

\begin{equation} 
X^{a}=i{1 \over m^{2}}
[{\bar m}_{(i)(j)}\omega^{A}_{(i)}\pi^{A^{\prime}}_{(j)}
-
m_{(i)(j)}{\bar \omega}^{A^{\prime}}_{(i)}{\bar \pi}^{A}_{(j)}].
\end{equation} 

\vskip 10pt

\noindent
Now it is a straightforward task to calculate the following
Poincar{\'e} invariant and Poincar{\'e} covariant Poisson bracket
commutation relations which will be needed in the sequel.

\vskip 10pt

\noindent
\it LEMMA 6: \rm

\vskip 10pt

\noindent
First we note that from the conformally invariant canonical Poisson
bracket relations (3.1) it follows that the $2n^{2}-n$ scalars in
(3.7)-(3.9) form a Poincar{\'e} invariant closed algebra of Poisson
brackets:

\begin{equation}
\{a_{(i)(j)}, \ a_{(k)(l)} \}=
ia_{(k)(j)}
\delta_{(i)(l)}-ia_{(i)(l)}\delta_{(j)(k)},
\ \ 
\{a_{(i)(j)}, \ m_{(k)(l)} \}=
2im_{(i)[(k)}\delta_{(l)](j)}
\end{equation}

\begin{equation}
\{a_{(i)(j)}, \ {\bar m}_{(k)(l)} \}=
2i{\bar m}_{(j) [(k) } \delta_{(l)](i) },
\ \ \ 
\{m_{(i)(j)}, \ m_{(k)(l)} \}=
\{{\bar m}_{(i)(j)}, \ m_{(k)(l)} \}=0.
\end{equation}

\vskip 10pt

\noindent
which may easily be proved by the help of lemma $4$.

\vskip 10pt

\noindent
\it LEMMA 7: \rm

\vskip 10pt

\noindent
From the fact that $m_{(i)(j)}$, ${\bar m}_{(i)(j)}$ and
$a_{(i)(j)}$ are Poincar{\'e} scalar functions it trivially
follows that they commute with all the generators of the Poincar{\'e}
algebra:

\begin{equation}
\{a_{(i)(j)}, \ {\cal P}_{a} \}=
\{a_{(i)(j)}, \ {\cal M}_{ab} \}=0
\end{equation}

\begin{equation}
\{m_{(i)(j)}, \ {\cal P}_{a} \}=
\{m_{(i)(j)}, \ {\cal M}_{ab} \}=
\{{\bar m}_{(i)(j)}, \ {\cal P}_{a} \}=
\{{\bar m}_{(i)(j)}, \ {\cal M}_{ab} \}=0.
\end{equation}

\vskip 10pt

\noindent
\it LEMMA 8: \rm

\vskip 10pt

\noindent
The following commutation relations are also easily deduced from the
canonical commutations relations in (3.1):

\begin{equation}
\{a_{(i)(j)}, l \}=0
\end{equation}

\begin{equation}
\{m_{(i)(j)}, \ l \}=m_{(i)(j)}.
\end{equation}

\vskip 10pt

\noindent
\it LEMMA 9: \rm

\vskip 10pt

\noindent
From (3.13) and  (3.17)-(3.22) it now follows that:

\begin{equation}
\{a_{(i)(j)}, \ X^{a} \}={1 \over m^{2}}{\cal P}^{a}
\{a_{(i)(j)}, l \}=0
\end{equation}

\begin{equation}
\{m_{(i)(j)}, \ X^{a} \}={1 \over m^{2}}{\cal P}^{a}
\{m_{(i)(j)}, \ l \}={1 \over m^{2}}{\cal P}^{a}m_{(i)(j)}
\end{equation}

\begin{equation}
\{{\cal P}_{b}, \ X^{a} \}=\delta^{a}_{b}
\end{equation}

\begin{equation}
\{{1 \over 2}{\cal P}_{b}{\cal P}^{b}, \ X^{a} \}={\cal P}^{a}.
\end{equation}

\noindent
\it LEMMA 10: \rm

\vskip 10pt

\noindent
Similarily we obtain that:

\begin{equation} 
\{X^{a}, \ X^{b}\}={1 \over m^{4}}\epsilon^{abcd}{\cal S}_{c}{\cal P}_{d}
\end{equation} 

\vskip 10pt

\noindent
where the Pauli-Luba{\'n}ski four-vector ${\cal S}^{a}$ is defined as
in (2.12) i.e.  by:

\begin{equation} 
{\cal S}^{a}:={1 \over 2}\epsilon^{abcd}{\cal P}_{b}{\cal M}_{cd}.
\end{equation} 

\vskip 10pt

\noindent
\it LEMMA 11: \rm

\vskip 10pt

\noindent
Expressing the right hand side of (3.28) 
in terms of spinors defining the corresponding twistors we obtain:

\begin{equation} 
2{\cal S}^{a}=2{\cal S}^{AA^{\prime}}=
{\bar m}_{(i)(j)}\omega^{A}_{(i)}\pi^{A^{\prime}}_{(j)}
+
m_{(i)(j)}{\bar \omega}^{A^{\prime}}_{(i)}{\bar \pi}^{A}_{(j)}
+
a_{(i)(j)}{\bar \pi}^{A}_{(i)}\pi^{A^{\prime}}_{(j)}
\end{equation} 

\vskip 10pt

\noindent
which after some spinor algebra manipulations may be rewritten as:

\begin{equation} 
2{\cal S}^{AA^{\prime}}=
[2a_{(i)(j)}-\delta_{(i)(j)}a_{(k)(k)}]
{\bar \pi}^{A}_{(i)}\pi^{A^{\prime}}_{(j)}=
2a_{(i)(j)}{\bar \pi}^{A}_{(i)}\pi^{A^{\prime}}_{(j)}-
a_{(i)(i)}
{\bar \pi}^{A}_{(j)}\pi^{A^{\prime}}_{(j)}.
\end{equation} 

\vskip 10pt

\noindent
\sl REMARK 8:\rm

\vskip 10pt

\noindent 
The formula in (3.30) is also valid for $n=1$ and reproduces the
result\cite{prmc} of lemma $2$.
For $n=2$ it appeared in Tod's
doctoral dissertation\cite{todd}.  However, the author of this paper
has never come across the general formula in (3.30) which is valid for
any (finite) natural number $n$.

\vskip 10pt

\noindent
\it LEMMA 12: \rm

\vskip 10pt

\noindent 
From the above lemma (lemma 11) it follows that the square of the
value of the total spin $s^{2}$ (for $n > 2$) is a function of the
invariants in (3.7)-(3.9) (i.e.  is a function of the generators of
the internal symmetries) given by:

$$-4m^{2}s^{2}:=4{\cal S}^{a}{\cal S}_{a}=$$
\begin{equation} 
(a_{(j)(j)})^{2}m^{2} + 4a_{(j)(j)}a_{(u)(v)}{\bar m}_{(u)(k)}m_{(k)(v)}
+4a_{(j)(k)}a_{(u)(v)}{\bar m}_{(j)(u)}m_{(k)(v)}.
\end{equation} 

\vskip 10pt

\noindent
\sl REMARK 9:\rm

\vskip 10pt

\noindent 
For $n=2$ and $n=3$ the formula (3.31) agrees with those previously
derived by Perj{\`e}s, Hughston, Sparling.  The general formula above
is, however, valid for any (finite) natural number $n \geq 2$.  As far
as we know this formula has not been derived before.

\vskip 10pt

\noindent
\bf PROPOSITION 1:\rm

\vskip 10pt

\noindent 
As explained in the introduction we wish to regard a closed massive
and spinning system as composed of a finite number of interacting
massless parts.  For this reason we notice that any function of the
form

\begin{equation} 
H:={1 \over 2}{\cal P}_{b}{\cal P}^{b} + 
g(a_{(j)(k)},  \ m_{(l)(m)}, \ {\bar m}_{(n)(r)}),
\end{equation} 

\vskip 10pt

\noindent
where $g$ is a positive real valued function of the invariants
in (3.7)-(3.9), generates a canonical flow in \bf Tp(n) \rm which
in the Minkowski space describes a set of $n$ mutually interacting
massless particles. This follows from direct calculations 
which produce the following equations of the motion:

\begin{equation} 
{\dot X}^{a}=
\{H, \ X^{a}\}=
{\cal P}^{a}(1+{1 \over m^{2}}
{{\partial g} \over {\partial m^{(i)(k)}}}m_{(i)(k)}
+{1 \over m^{2}}
{{\partial g} \over {\partial {\bar m}^{(i)(k)}}}{\bar m}_{(i)(k)}
),
\end{equation} 

\begin{equation} 
{\dot \pi}_{A^\prime (k)} = \{H, \ \pi_{A^\prime (k)} \}=
-i{{\partial g} \over {\partial a^{(j)(k)}}}\pi_{A^\prime (j)},
\qquad \qquad
{\dot {\cal P}}^{a}=
\{H, \ {\cal P}^{a}\}=0,
\end{equation} 

$${\dot a}_{(k)(j)}=\{H, \ a_{(k)(j)} \}=$$
\begin{equation} 
i{{\partial g} \over {\partial a^{(j)(l)}}}a_{(k)(l)}
-
i{{\partial g} \over {\partial a^{(l)(k)}}}a_{(l)(j)}
-
2i{{\partial g} \over {\partial {\bar m}^{(k)(l)}}}{\bar m}_{(l)(j)}
+
2i{{\partial g} \over {\partial m^{(j)(l)}}}m_{(k)(l)}.
\end{equation} 

\vskip 10pt

\noindent
\sl REMARK 10:\rm

\vskip 10pt

\noindent 
Note that the assumption in (3.32), stating that $g$ is a
function of the generators $a_{(i)(j)}$ (which are conformal scalars),
makes the motion of the massless parts non-trivial (i.e.  changes
their null-momenta during the motion).  All functions $g$, which
depend on $m_{(i)(j)}$ and their complex conjugates only, produce a
motion of the massless parts which is trivial in the Minkowski space.

\vskip 10pt

\noindent
\it ASSUMPTION 1:\rm

\vskip 10pt

\noindent
From now on we assume that $g$ is a function of the conformal invariants 
$a_{(i)(j)}$ only:

\begin{equation} 
g=g(a_{(j)(k)}). 
\end{equation} 

\vskip 10pt

\noindent
\bf PROPOSITION 2:\rm

\vskip 10pt

\noindent
Under this condition the equations of motion generated by the canonical flow
in the twistor phase space simplify and read:

\begin{equation} 
{\dot X}^{a}=
\{H, \ X^{a}\}={\cal P}^{a},
\end{equation} 

\begin{equation} 
{\dot \pi}_{A^\prime (k)} = \{H, \ \pi_{A^\prime (k)} \}=
-i{{\partial g} \over {\partial a^{(j)(k)}}}\pi_{A^\prime (j)},
\qquad \qquad
{\dot {\cal P}}^{a}=
\{H, \ {\cal P}^{a}\}=0,
\end{equation} 

\begin{equation} 
{\dot a}_{(k)(j)}=\{H, \ a_{(k)(j)} \}=
i{{\partial g} \over {\partial a^{(j)(l)}}}a_{(k)(l)}
-
i{{\partial g} \over {\partial a^{(l)(k)}}}a_{(l)(j)}.
\end{equation} 

\vskip 10pt

\noindent
\sl REMARK 11:\rm

\vskip 10pt

\noindent
From (3.37) it follows that the parameter labelling points on the
curves of the canonical flow generated by $H$ with $g$ such as in
(3.36) is linearly related to the proper time of the total system.

\vskip 10pt

\noindent
If such a function $g$ vanishes (or degenerates to a real number) then
the function $H$ and the function ${1 \over 2} m^{2}$ are identical
(modulo an additive real number) forming just one constant of the free motion
generated by $H$.

\vskip 10pt

\noindent
For non-trivial such $g$ the functions $H$ and ${1 \over 2} m^{2}$
correspond to two different mutually commuting but {\`a} priori
unrelated constants of the motion generated by $H$.

\vskip 10pt

\noindent
\sl REMARK 12:\rm

\vskip 10pt

\noindent 
The above equations of motion have been explicitly solved\cite{ab1}
for $n=2$ and for $g=s^{2}$.  Such a motion describes a massive
relativistic rigid rotator composed of two directly interacting
massless spinning particles.

\vskip 10pt

\noindent
\it ASSUMPTION 2:\rm

\vskip 10pt

\noindent
Due to the fact that the parameter labelling the curves of the
canonical flow generated by $H$ is linearly related to the proper time
of the total system, we make an additional "physical" assumption that
for any non-trivial $g$ such as in (3.36) the constant of the motion
given by the value of the function $H$ is proportional (modulo an
additive real number $r$) to the value of the constant of the motion
${1 \over 2}m^{2}$.  The proportionality constant $k$ is larger than
one half and aproaches one half when the function $g$ approaches zero
(modulo an additive real number $r$):

\begin{equation} 
H=km^{2} + r \ \ \ \ \ \ \ \ \ k > {1 \over 2}
\end{equation} 

\vskip 10pt

\noindent
{\`A} posteriori this amounts to a "constraint":

\begin{equation} 
m^{2}={g - r \over (k-{1 \over 2})}
\end{equation} 

\vskip 10pt

\noindent
The imposition of such a "constraint" seems perhaps somewhat
unnecessary at this stage but may be motivated by the fact that after
quantization we wish to interpret ratios of the arising possibly
discrete eigenvalues of ${\hat H}$ (for some specific choices of ${g}$
in (3.36)) as ratios of the squares of the quantized masses.

\section{\hspace{-4mm}\hspace{2mm}QUANTIZATION.}

A non-standard\cite{woo}, as opposed to the standard procedure
introduced by Penrose\cite{prmc}, canonical twistor quantization is
obtained by means of a natural prescription {\'a} la
Dirac\cite{dpam1,dpam2} given by:

\begin{equation}{\hat \omega}^{A}_{(i)} :=  
- {{\partial} \over {\partial {\bar \pi}_{A}^{(i)}}}, \qquad \qquad 
{\hat {\bar \omega}}^{A^\prime}_{(i)} := {{\partial} \over {\partial
{\pi}_{A^\prime}^{(i)}}},\end{equation}

\begin{equation}{\hat {\bar \pi}}_{A(i)} := {\bar \pi}_{A(i)}, \qquad \qquad
{\hat {\pi}}_{A^\prime (i)} := {\pi}_{A^\prime (i)}.\end{equation}

\vskip 10pt

\noindent
The Poisson brackets relations in (3.1) will hereby be
replaced by the corresponding commutators turning the classical
twistor phase space dynamics of massless particles into its quantum
mechanical analog.

\vskip 10pt

\noindent
So by the use of (4.1)-(4.2) the linear four-momentum functions in
(3.2), the angular four-momentum functions in (3.3), the scalar
functions in (3.7)-(3.9) turn into the corresponding operators:

\begin{equation}{\hat {\cal P}}_{a}:={\bar \pi}_{A(i)}{\pi}_{A^\prime (i)},\end{equation}

\begin{equation} {\hat {\cal M}}^{ab}:=
i{\pi}^{({A^\prime}}_{(i)}
{{\partial} \over {\partial
{\pi}_{{B^\prime})}^{(i)}
}}
\epsilon^{AB}
+
i{\bar \pi}^{({A}}_{(i)}
{{\partial} \over {\partial {\bar \pi}_{{B)}}^{(i)}
}}
\epsilon^{{A^\prime}{B^\prime}},
\end{equation}

\begin{equation}{\hat a_{(i)(j)}}: =  
-
{\bar \pi}_{A(j)}{\partial \over \partial {\bar \pi}_{A}^{(i)}} +
\pi_{A^\prime (i)}{\partial \over \partial \pi_{A^\prime}^{(j)}},
\ \ \
{\hat m_{(i)(j)}}: =  {\pi}^{A^\prime}_{(i)}{\pi}_{A^\prime (j)},
\ \ \ 
{\hat {\bar m}_{(i)(j)}}: = {\bar \pi}^{A}_{(i)}{\bar \pi}_{A(j)}.
\end{equation}

\vskip 10pt

\noindent
The Poisson bracket relations in (3.4)-(3.6) ensure that operators in
(4.3) and (4.4) obey commutation relations of the Poincar{\'e}
algebra.  In addition all the Poisson bracket commutation relations 
in (3.10) - (3.12), (3.17) - (3.27) turn into the corresponding
operator commutators. 

\vskip 10pt

\noindent
All functions on \bf Tp(n) \rm become (at least formally) operator
valued functions of the canonical differential operators in (4.1) and
of the multiplicative operators in (4.2).
Of course this may lead to problems: Ordering problems,
non-locality of the operators arising from functions on \bf Tp(n) \rm
in which the "$\omega$" parts appear in the denominator etc.

\vskip 10pt

\noindent
The multiplicative operators in (4.2) define a Poincar{\'e} invariant
$4n$ real dimensional configuration vector space $\Pi$ spanned by $n$
Weyl spinors and their complex conjugates.

\vskip 10pt

\noindent
The above (formally) defined operator valued functions of the
canonical operators in (4.1)-(4.2) act on the infinitely dimensional
space $\Gamma$ of complex valued "wave" functions defined on $\Pi$.

\vskip 10pt

\noindent
A Poincar{\'e} invariant scalar product on the space of complex valued
functions on $\Pi$ we tentatively define as: 

\begin{equation}<f_{1}\vert f_{2}>:=
\int
[{\bar f}_{1}({\bar \pi}_{B (i)}, \ \pi_{B^\prime (i)})
f_{2}(\pi_{B^\prime (i)}, \ {\bar \pi}_{B (i)})]
d\pi^{A^\prime}_{(j)} \wedge
d\pi_{A^\prime (j)} \wedge d{\bar \pi}^{A}_{(k)} \wedge 
d{\bar \pi}_{A (k)}.\end{equation}

\vskip 10pt

\noindent
The subspace ${\aleph}$ of $\Gamma$ consisting of functions having
finite norms with respect to this scalar product defines a Hilbert
space of quantum states of the massive spinning composite particle.

\vskip 10pt

\noindent
The quantized version of our general dynamical principle now reads:

\vskip 10pt

\noindent
Find common eigenvalues and eigenfunctions of a maximal set of
hermitian mutually commuting operators containing the following subset
($N$ refers to the normal ordering of terms):

\begin{equation}
{\hat {\cal P}}_{a}={\bar \pi}_{A(i)}{\pi}_{A^\prime (i)},
\end{equation}

\begin{equation} 
{\hat H}={1 \over 2}{\hat m}^{2}+N(g({\hat a}_{(i)(j)}))
\end{equation} 

\vskip 10pt

$$S_{z}=-{1 \over m}{\hat {\cal S}}^{a}n_{a}=
-{1 \over m}{\hat {\cal S}}^{AA^{\prime}}n_{AA^\prime}=$$
\begin{equation}
{1 \over 2}({\hat a}_{(i)(j)}
{\bar \pi}^{A}_{(i)}\pi^{A^{\prime}}_{(j)}+
{\bar \pi}^{A}_{(i)}\pi^{A^{\prime}}_{(j)}
{\hat a}_{(i)(j)})n_{AA^\prime}
-
{1 \over 4}({\hat a}_{(i)(i)}
{\bar \pi}^{A}_{(j)}\pi^{A^{\prime}}_{(j)}+
{\bar \pi}^{A}_{(j)}\pi^{A^{\prime}}_{(j)}
{\hat a}_{(i)(i)})n_{AA^\prime}.
\end{equation} 

\vskip 10pt

$${\hat m}^{2}{\hat s}^{2}:=-{\hat {\cal S}}^{a}{\hat {\cal S}}_{a}=$$
\begin{equation} 
-{1 \over 4}({\hat a}_{(j)(j)})^{2}m^{2} - 
N({\hat a}_{(j)(j)}{\hat a}_{(u)(v)}{\bar m}_{(u)(k)}m_{(k)(v)})
-
N({\hat a}_{(j)(k)}{\hat a}_{(u)(v)}{\bar m}_{(j)(u)}m_{(k)(v)}).
\end{equation} 

\vskip 10pt

\noindent
where $m$ denotes an eigenvalue of the mass and $n_{a}$ is any
space-like unit four-vector orthogonal to the time-like direction
defined by ${\hat {\cal P}}_{a}$.  In other words $n^{a}$ represents a
"z-axis" direction.

\vskip 10pt

\noindent
From the assumption $2$ and general group theoretical considerations
it follows that eigenvalues of the mass squared ${\hat m}^{2}$ are
proportional (up to an additive real number) to the eigenvalues of
$N(g({\hat a}_{(i)(j)}))$ (which is assumed to be hermitian). Moreover
the eigenvalues of the square of the spin ${\hat s}^{2}$ and of ${\hat
S}_{z}$ assume the usual values i.e. $j(j+1)$ and $j_{z}=-j, ...., j$
with $j$ being a positive integral number or a positive half integral
number.

\vskip 10pt

\noindent 
For each choice of $g$ one can choose among the operators in (4.5) a
maximal set of mutually commuting ones which also commute with
$N(g({\hat a}_{(i)(j)}))$.  The eigenvalues of these additional
operators may be identified with the internal degrees of freedom of
the total system.  To each set of eigenvalues of the mutually
commuting internal operators and to each set of the mutually commuting
external operators' eigenvalues (the mass, $j$, $j_{z}$, total
four-momentum of the system) there corresponds a state function in
$\aleph$ which may be calculated (at least in principle) using methods
from non-relativistic quantum mechanics.

\vskip 10pt 

\noindent
As an example consider the relativistic rigid rotator composed of two
($n=2$) massless constituents with helicity\cite{ab1}.  To quantize it
we note that the external commuting observables may
be chosen as:

\begin{equation}
{\hat {\cal P}}_{a}={\bar \pi}_{A(1)}{\pi}_{A^\prime (1)}
+{\bar \pi}_{A(2)}{\pi}_{A^\prime (2)},
\end{equation}

\begin{equation} 
{\hat H}={1 \over 2}{\hat m}^{2}+{\hat s}^{2},
\end{equation} 

$${\hat {\cal S}}^{a}n_{a}=
{\cal S}^{AA^{\prime}}n_{AA^\prime}=$$
$${1 \over 4}({\hat a}_{(1)(1)}-{\hat a}_{(2)(2)})
({\bar \pi}^{A}_{(1)}\pi^{A^{\prime}}_{(1)}
-{\bar \pi}^{A}_{(2)}\pi^{A^{\prime}}_{(2)})n_{AA^\prime}
+
({\bar \pi}^{A}_{(1)}\pi^{A^{\prime}}_{(1)}
-{\bar \pi}^{A}_{(2)}\pi^{A^{\prime}}_{(2)})n_{AA^\prime}
{1 \over 4}({\hat a}_{(1)(1)}-{\hat a}_{(2)(2)})$$
\begin{equation}
+
{1 \over 2}({\hat a}_{(1)(2)}
{\bar \pi}^{A}_{(1)}\pi^{A^{\prime}}_{(2)}+
{\bar \pi}^{A}_{(1)}\pi^{A^{\prime}}_{(2)}
{\hat a}_{(1)(2)})n_{AA^\prime} +
{1 \over 2}({\hat a}_{(2)(1)}
{\bar \pi}^{A}_{(2)}\pi^{A^{\prime}}_{(1)}+
{\bar \pi}^{A}_{(2)}\pi^{A^{\prime}}_{(1)}
{\hat a}_{(2)(1)})n_{AA^\prime},
\end{equation} 

\begin{equation} 
{\hat s}^{2}={1 \over 4}
({\hat a}_{(1)(1)}-{\hat a}_{(2)(2)})^{2}+
{1 \over 2}({\hat a}_{(1)(2)}{\hat a}_{(2)(1)}+
{\hat a}_{(2)(1)}{\hat a}_{(1)(2)}),
\end{equation} 

\vskip 10pt

\noindent
while internal symmetry operators are given e.g. by:

\begin{equation} 
{\hat a}_{(1)(1)} \qquad \qquad
{\hat a}_{(2)(2)}.
\end{equation} 

\vskip 10pt

\noindent
In the rigid rotator case the eigenvalues of the square of the mass are
proportional to $j(j+1)$ i.e.  proportional to the eigenvalues of the
square of the spin.  In addition the states (eigenfunctions) of the
rigid rotator are labelled by the eigenvalues of the Euler operators
in (4.15).

\vskip 10pt

\noindent
To find these relativistic rigid rotator eigenfunctions
$f(\pi^{A^{\prime}}_{(1)}, \
\pi^{A^{\prime}}_{(2)}, \ {\bar \pi}^{A}_{(1)}, \ {\bar
\pi}^{A}_{(2)})$ in $\aleph$ is a
much harder task and will not be pursuited in this paper.

\section{\hspace{-4mm}\hspace{2mm}CONCLUSIONS AND REMARKS.} 

If "elementary" particles such as e.g.  electron, proton, neutron etc.
may be regarded as bound states of a {\underline {finite}} number of
massless spinning parts then twistor theory combined with the idea of
relativistic action at a distance provide a very powerful tool for
construction of such models.

\vskip 10pt

\noindent
In such approaches, as shown in this paper, particle aspects of
Penrose's twistor formalism should be emphasized as opposed to the
standard treatments where field aspects are at the front.

\vskip 10pt

\noindent
The non-standard quantization procedure in (4.1) - (4.2) implies that
we loose some of the results of conventional twistor theory such as
the twistor description of massless free fields in terms of
holomorphic sheaf cohomology, the scalar product on such fields,
geometrization of the concept of positive frequency of the field and
the relationship between conformal curvature and the twistor
"position" (twistor variables) and "momentum" (complex conjugates of
the twistor variables) operators\cite{pr3}.

\vskip 10pt

\noindent
What we gain is that the real dimension of the relativistic
configuration space of massless spinning particles is one half of the
real dimension of the configuration space obtained by means of the
conventional holomorphic twistor quantization\cite{prmc}.  Further,
the configuration space obtained in our paper may be given a clear
physical interpretation.  Wave functions on such a configuration space
define quantum states in (the "square root" of) the linear
four-momentum representation of the massless parts.  However in our
opinion the most important gain is the fact that using our formulation
we are able to treat \it interacting \rm massless spinning particles
(not fields) forming a closed composite bound system.

\vskip 10pt

\noindent
To apply our ideas to concrete physical systems is, at the moment,
hampered by the fact that there are, as yet, no indications in the
model how the function $g$ in (3.36) should be chosen.

\vskip 10pt 

\noindent
Nevertheless, the general principle presented in this paper seems to
comply with the Twistor Programme announced by Penrose\cite{penr}.

\section*{Acknowledgment.}

Certain features of the idea presented in this note have been previously
described in a talk given at VII-th Conference on "Symmetry Methods in
Physics" - Dubna, 10-16 July, 1995.  The visit to this conference was
partly supported by NFR - Sweden, contract number R-RA 04666-306.

\end{document}